\documentclass[aps,prl,twocolumn,superscriptaddress]{revtex4-1}

\usepackage[pdftex]{graphicx}

\begin{document}

\title{Effect of tensile stress on the in-plane resistivity anisotropy in BaFe$_2$As$_2$ }%

\author{E.~C.~Blomberg}
\affiliation{The Ames Laboratory, Ames, Iowa 50011, USA}
\affiliation{Department of Physics and Astronomy, Iowa State University, Ames, Iowa 50011, USA }

\author{A.~Kreyssig}
\affiliation{The Ames Laboratory, Ames, Iowa 50011, USA}
\affiliation{Department of Physics and Astronomy, Iowa State University, Ames, Iowa 50011, USA }

\author{M.~A.~Tanatar}
\affiliation{The Ames Laboratory, Ames, Iowa 50011, USA}

\author{R.~Fernandes}
\altaffiliation[Current address: ]{Department of Physics, Columbia University, New York, New York 10027, USA}
\affiliation{The Ames Laboratory, Ames, Iowa 50011, USA}
\affiliation{Department of Physics and Astronomy, Iowa State University, Ames, Iowa 50011, USA }

\author{M.~G.~Kim}
\affiliation{The Ames Laboratory, Ames, Iowa 50011, USA}
\affiliation{Department of Physics and Astronomy, Iowa State University, Ames, Iowa 50011, USA }

\author{A.~Thaler}
\affiliation{The Ames Laboratory, Ames, Iowa 50011, USA}
\affiliation{Department of Physics and Astronomy, Iowa State University, Ames, Iowa 50011,
USA }

\author{J.~Schmalian}
\affiliation{The Ames Laboratory, Ames, Iowa 50011, USA}
\affiliation{Department of Physics and Astronomy, Iowa State University, Ames, Iowa 50011,
USA }

\author{S.~L.~Bud'ko}
\affiliation{The Ames Laboratory, Ames, Iowa 50011, USA}
\affiliation{Department of Physics and Astronomy, Iowa State University, Ames, Iowa 50011,
USA }
\author{P.~C.~Canfield}
\affiliation{The Ames Laboratory, Ames, Iowa 50011, USA}
\affiliation{Department of Physics and Astronomy, Iowa State University, Ames, Iowa 50011,
USA }
\author{A.~I.~Goldman}
\affiliation{The Ames Laboratory, Ames, Iowa 50011, USA}
\affiliation{Department of Physics and Astronomy, Iowa State University, Ames, Iowa 50011,
USA }

\author{R.~Prozorov}
\email[Corresponding author: ]{prozorov@ameslab.gov}
\affiliation{The Ames Laboratory, Ames, Iowa 50011, USA}
\affiliation{Department of Physics and Astronomy, Iowa State University, Ames, Iowa 50011,
USA }

\date{31 October 2011}

\begin{abstract}

The effect of uniaxial tensile stress and the resultant strain on the structural/magnetic transition in the parent compound of the iron arsenide superconductor, BaFe$_2$As$_2$, is characterized by temperature-dependent electrical resistivity, x-ray diffraction and quantitative polarized light imaging. We show that strain induces a measurable uniaxial structural distortion above the first-order magnetic transition and significantly smears the structural transition. This response is different from that found in another parent compound, SrFe$_2$As$_2$, where the coupled structural and magnetic transitions are strongly first order. This difference in the structural responses explains the in-plain resistivity anisotropy above the transition in BaFe$_2$As$_2$. This conclusion is supported by the Ginzburg-Landau - type phenomenological model for the effect of the uniaxial strain on the resistivity anisotropy.

\end{abstract}

\maketitle

\section{Introduction}

At ambient conditions, the parent compounds of iron-arsenide superconductors, $A$Fe$_2$As$_2$ ($A$ = Ba, Ca or Sr), crystallize in the tetragonal ThCr$_2$Si$_2$ structure \cite{Rotter,Ca-phasetransition}. On cooling, they undergo a structural phase transition with the lattice symmetry lowered from tetragonal to orthorhombic at a characteristic temperature $T_{TO}$ (170~K for $A$=Ca \cite{Rotter}, \cite{Ca-phasetransition}, 205~K for $A$=Sr \cite{Yan} and 135~K for $A$=Ba \cite{Rotter}. We denote the compounds as A122 in the following). This transition is accompanied or followed by long-range magnetic ordering into an antiferromagnetic (AFM) stripe phase at the Ne\'el temperature, $T_N$ \cite{Rotterneutrons}. Indeed $T_N = T_{TO}$ in compounds with $A$=Ca \cite{Caneutrons} and $A$=Sr \cite{Srneutrons}, where the transition is sharp and strongly first order. In BaFe$_2$As$_2$ $T_N \le T_{TO}$, and the structural transition is second order, whereas the AFM transition is first order \cite{Birgeneau,Kreyssigsplit}.

Since the doping (or pressure) - dependent superconductivity in A122 iron arsenides exhibits the  highest $T_c$ close to the point of complete suppression of the structural/magnetic order, understanding the mechanism of these transitions is very important for understanding the nature of superconductivity. The parent compounds of iron arsenides are metals, so it is suggested that their magnetism is of itinerant character due to a spin density wave (SDW) instability of the multi-band Fermi surface \cite{MazinSDW,ChubukovSDW}. On the other hand, it has also been suggested that the magnetism can arise in a local moment picture due to magnetic frustration \cite{Abrahams} and/or orbital ordering \cite{orbital1,orbital2,orbital3,orbital4}. Therefore, it is important to conduct measurements that characterize the normal state anisotropy of the electronic structure in the vicinity of $T_N$. From first principles calculations, the electronic anisotropy of iron pnictides was predicted to be fairly high in the orthorhombic $\bf{ab}-$plane below $T_N$ \cite{detwinning1,1stprinciple1,1stprinciple2,1stprinciple3,1stprinciple4}. On the other hand, ARPES measurements suggest that a notable energy splitting between $d_{xz}$ and $d_{yz}$ orbitals appears below the transition \cite{Shen,Korean}.

An insight into intrinsic anisotropy became possible after the development of detwinning techniques, using uniaxial tensile \cite{detwinning1,detwinning2} or compressive stress \cite{Fisher1,Fisher2} (see Ref.~\onlinecite{Fisher-review} for a review). Electrical resistivity measurements in the detwinned state found the in-plane resistivity anisotropy to have an unusual temperature dependence of the ratio $\rho_b/\rho_a$, peaking just below $T_{TO}$ with maximum $\rho_b/\rho_a=1.2, 1.4, 1.5$ for $A$=Ca, Sr, Ba, respectively \cite{detwinning1,detwinning2}. Surprisingly, in BaFe$_2$As$_2$ the high temperature ``tail'' of the anisotropy is found even in the nominally tetragonal phase above $T_{TO}$. The anisotropy above $T_{TO}$ becomes most pronounced in slightly doped Ba(Fe$_{1-x}$Co$_x$)$_2$As$_2$ (BaCo122), $x<0.06$\cite{Fisher1}, but is not observed in either Ca122 or Sr122.
This difference between Ca122 and Sr122 compounds on one hand, and Ba122 and BaCo122 compounds on the other, may be related to the sharpness of the structural phase transition, i.e., a strongly coupled first order \cite{Caneutrons}, \cite{Srneutrons} vs. split second order transitions \cite{Birgeneau, Kreyssigsplit}.

In this paper, we study the evolution of the electronic and structural anisotropy of detwinned Ba122 with special attention to the effects of the applied strain required to detwin the samples. With a much better strain control, we find that the effect above the transition arises from an anomalously large structural susceptibility of the crystals to the applied strain. This strain further separates the already split structural and magnetic transitions in BaFe$_2$As$_2$, as found in the detailed analysis of the temperature-dependent x-ray and polarized optical imaging. We model the effect of the applied strain by using a phenomenological Ginzburg-Landau - type model and show that the difference in the response is directly linked to the order of the magnetic/structural transitions.

\begin{figure}[tb]
\begin{center}
\includegraphics[width=0.85\linewidth]{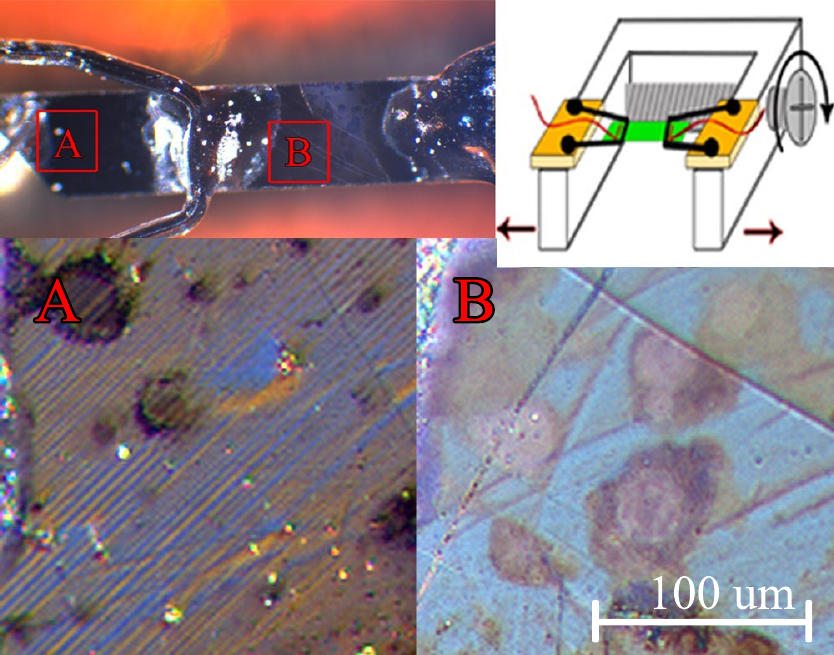}
\end{center}
\caption{(Color Online) The top right panel shows a schematic of the horseshoe with the potential leads used to apply the strain by adjusting the push-screw. The current leads are mounted strain-free. The top left panel shows the photograph of an actual sample with soldered contacts. The area ``A'' on the left side of the sample represents the unstrained region. Its polarized light image at 5 K (bottom left panel) reveals clear domain pattern by the alternating blue and orange stripes. The region ``B''  is located between the potential contacts in the strained part of the sample. It is shown in its detwinned state in the bottom right panel.}%
\label{1-detwin}%
\end{figure}

\begin{figure}[tb]
\begin{center}
\includegraphics[width=0.85\linewidth]{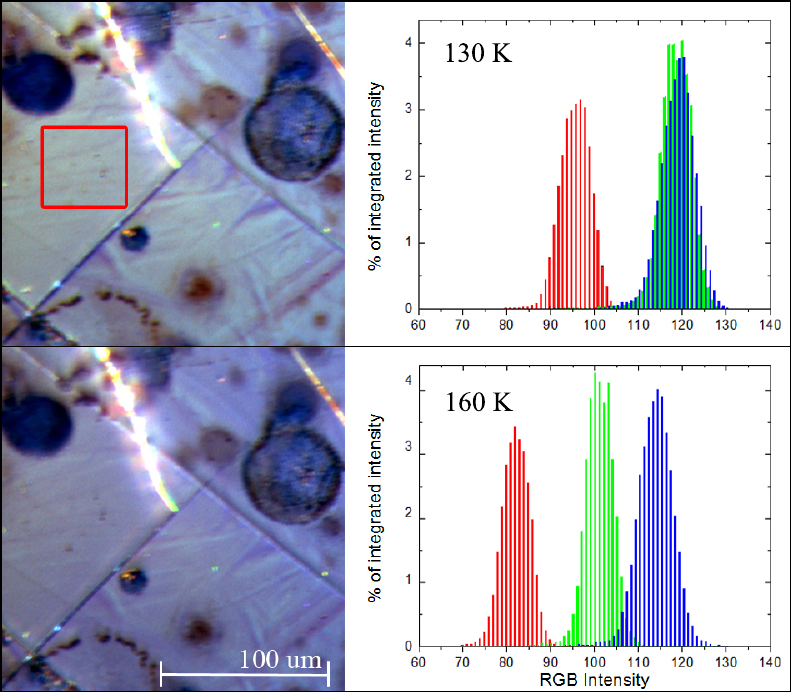}
\end{center}
\caption{(Color Online) Left panels show polarized light images of the strained portion ``B''  of the sample below (130~K, top) and above (160~K, bottom) the structural transition temperature. The intensity in red, green and blue (RGB) channels for each pixel was digitized using 256 intensity bins. The total intensity was found by summing the intensities of all pixels in the selected region. The RGB color was characterized by the percent contribution of each channel to the total intensity, plotted against the value of each bin to produce the histograms in Fig.~\ref{f2-RGB}. Right panels show the RGB histograms of a small area of the strained portion ``B''  of the sample, indicated by the red square in the top left panel. Whereas the blue channel remains almost unchanged, the intensities of the green and red channels shift dramatically indicating overall spectral change. The temperature dependence of this change is quantitatively analyzed in Fig.~\ref{3-color}. }%
\label{f2-RGB}%
\end{figure}

\section{Experimental Methods}

Single crystals of Ba122 were grown from an FeAs flux as described in Ref.~\onlinecite{Bacrystalgrowth}. Crystals were cut into strips with typical dimensions of 0.6 mm wide, 3 to 5 mm long and had a thickness of approximately 0.1 mm. The lengthwise cuts were made parallel to the tetragonal [110] direction, which will become the orthorhombic $\bf{a_o}$ or $\bf{b_o}$ axes below the transition temperature. Cutting directions were estimated by eye using polarized optical images of the domain structure and natural facets on the sample.

Polarized light images were taken at temperatures down to 5~K using a \textit{Leica DMLM} polarization microscope equipped with a flow-type $^{4}$He cryostat, as described in detail in Ref.~\onlinecite{domains}. High-resolution images were recorded with a spatial resolution of about 1 $\mu$m. Measurements were done with the polarizer and analyzer nearly crossed.

The leads for electrical resistivity measurements were formed by Ag wires, 50 $\mu$m in diameter, soldered to the samples with tin \cite{SUST}. A photograph of the sample with wires is shown in the top left panel in Fig.~\ref{1-detwin}. Four - probe measurements were conducted in a \textit{Quantum Design PPMS} from 5~K to 300~K. Measurements were first made on a free standing sample, and then the voltage leads were attached to a horseshoe, as schematically shown in the top-right panel of Fig.~\ref{1-detwin}, while the current leads were mounted so as to produce no strain on the sample. The strain was applied by means of a stainless push screw expanding the legs of the horseshoe. The temperature-dependent resistivity was measured after every strain increment. For the evaluation of the tensile stress value we compared our data with the data of T. Liang {\it et. al.} \cite{Liang}, who found a roughly +5 K shift of the tetragonal-to-orthorhombic transition feature upon a stress change from 15 to 50 MPa. In our case a total shift of approximately 3 K was achieved in four equivalent stress increments, which suggests that each strain increment is in the 4-5 MPa range for our horseshoe straining device. Therefore the strain at the highest level is estimated to be in the range of 16-20 MPa.

The sample was periodically imaged via polarized microscopy. The bottom panels in Fig.~\ref{1-detwin} show two areas of the sample: area $A$ (left panel) is located between current and potential leads and remains twinned during measurements; area $B$ (right panel) is located between the potential contacts in the strained part of the sample and becomes nearly twin-free under strain.
The application of uniaxial stress makes it energetically favorable to align domains with their long $\bf{a_o}$ axis along the strain, giving rise to an increasing volume fraction of one domain orientation above the rest.

\begin{figure}[tb]
\includegraphics[width=0.85\linewidth]{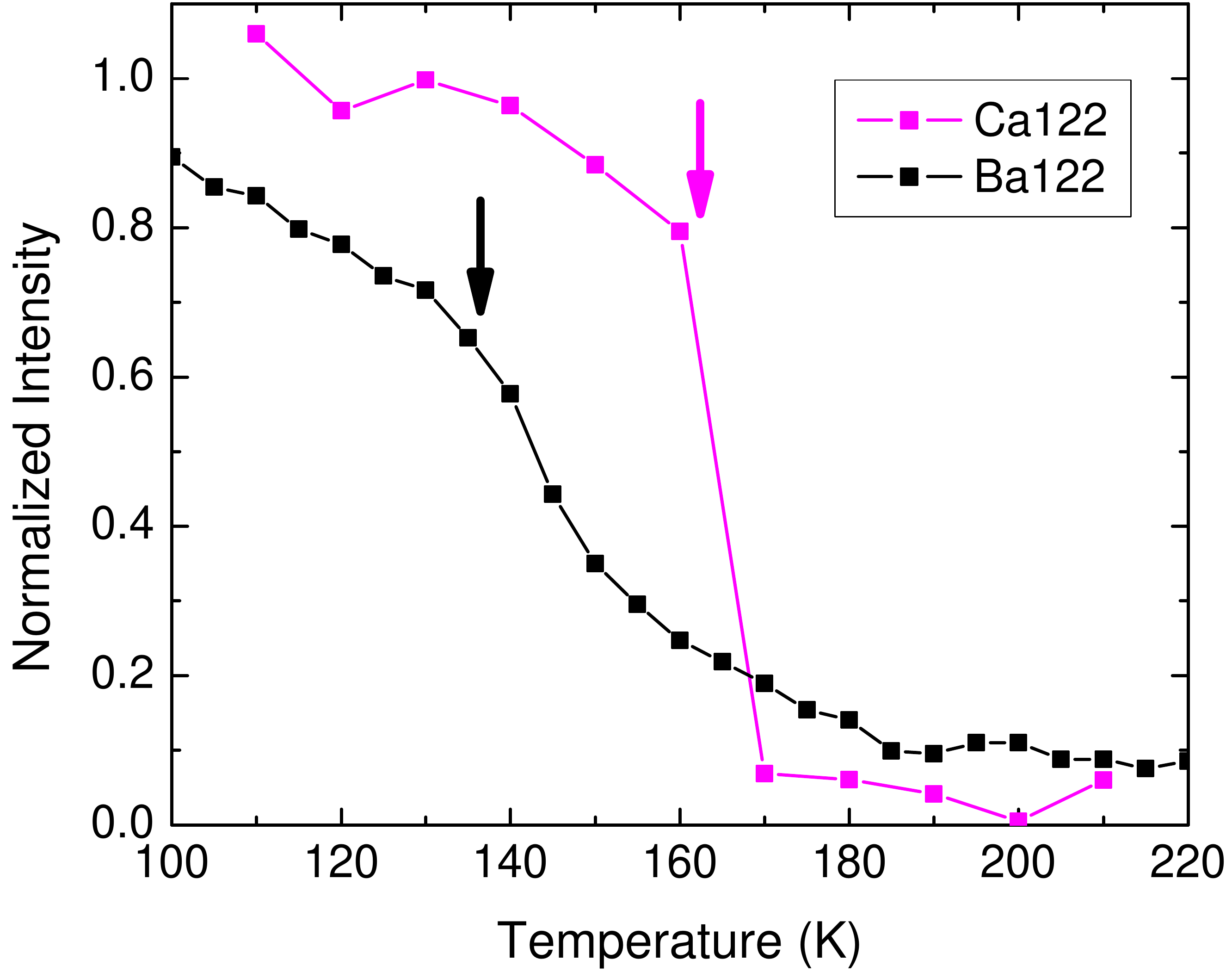}
\caption{(Color Online) Normalized temperature variation of the green color channel's intensity through the magnetic transition in detwinned crystals of  BaFe$_2$As$_2$ and  CaFe$_2$As$_2$. Arrows indicate the temperatures of the respective magnetic transitions. While there is a clear signature of the transition at about 135 K in BaFe$_2$As$_2$, the curve changes smoothly through the transition with a second order character. In CaFe$_2$As$_2$, the transition is quite sharp around 160 K, and is strongly first order. }
\label{3-color}
\end{figure}

The detwinned crystals were studied by high-energy x-ray diffraction in the MU-CAT sector (beamline 6ID-6) at the Advanced Photon Source at Argonne National Laboratory. Measurements were made with X-rays at 99.3 keV, giving an absorption length of approximately 1.5 mm and thus  allowing full penetration of the typically 0.1 mm thick samples. Entire reciprocal planes were recorded with the $\bf{c}$-axis parallel to the incident beam. During measurements, the samples were rocked through two independent angles perpendicular to the beam (see Ref. ~\onlinecite{kreyssigsetup}). The direct beam was blocked by a beam stop and diffraction images were recorded by a MAR345 image plate positioned 1680 mm behind the sample. The beam size was reduced to 0.2x0.2mm$^2$ by a slit system.

\begin{figure}[tb]
\begin{center}
\includegraphics[width=0.7\linewidth]{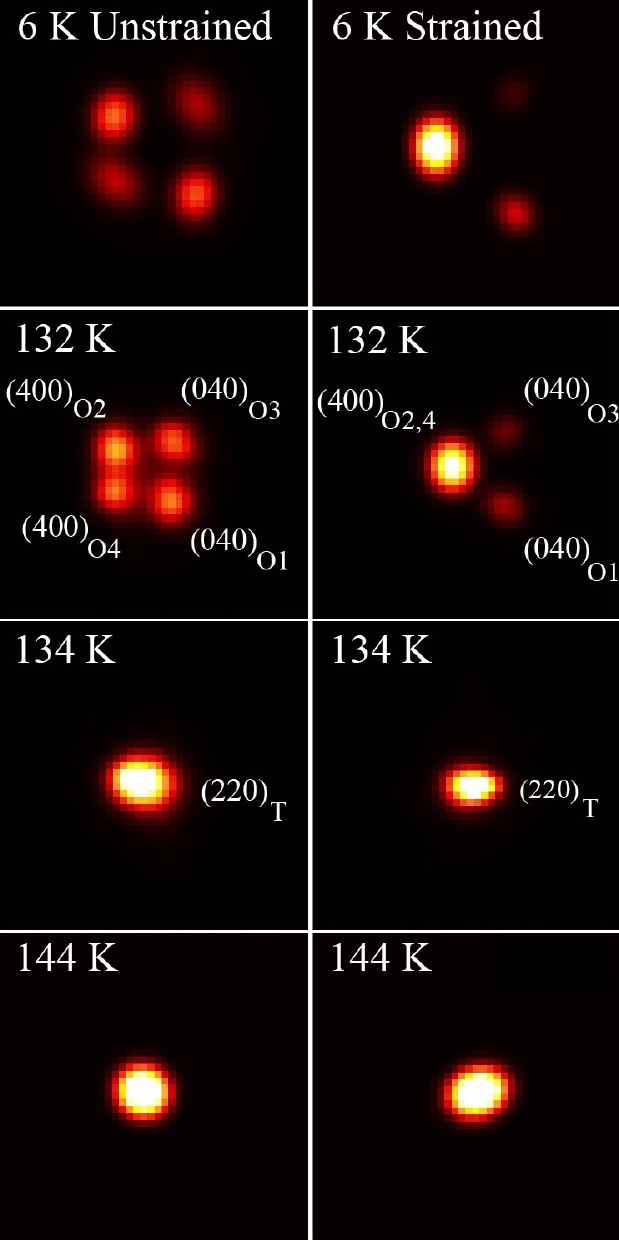}
\end{center}
\caption{ (Color Online) Temperature evolution of the two-dimensional x-ray diffraction pattern near the tetragonal (220) Bragg diffraction peak. Left and right columns of images show diffraction patterns in the unstrained and strained parts of the crystal, respectively. Four spots in the unstrained part at 6~K (top left) represent four domains in the sample with domain populations (proportional to integrated intensity) ranging between 19 and 31\%, close to random (see also Ref.~\onlinecite{domains}). In the strained portion of the sample (6 K, top right panel), the dominant domain occupies nearly 90 percent of the volume of the sample area probed by the x-ray beam. Between 132 and 134 K, the sample undergoes an orthorhombic to tetragonal structural transition. The second-order nature of the transition is evidenced by the lack of coexistence of orthorhombic and tetragonal domains. This coexistence is clearly observed in Sr122 (see Ref.~\onlinecite{detwinning2}), in which the transition is strongly first-order, see Fig.~\ref{5-schematics} below for schematic elaboration.}
\label{4-bragg}%
\end{figure}

\begin{figure}[tb]
\begin{center}
\includegraphics[width=1\linewidth]{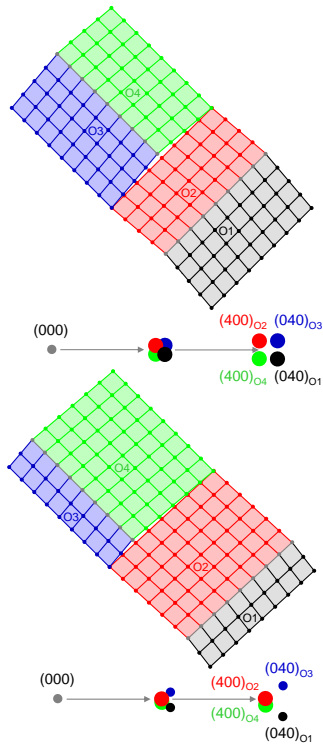}
\end{center}
\caption{ (Color Online) Schematic diagrams of the displacements of atoms in the twinned orthorhombic phase and the resulting Bragg reflections. As demonstrated in the top panel, a perfect crystal with equal populations of each domain orientation results in a square pattern between the (400) and (040) orthorhombic reflections. Conversely, the bottom panel illustrates the result of an unequal distribution of domain orientations. Here the angle between the O4 and the O2 domain orientations is significantly smaller and consequently moves the reflections closer together. Further, the population of each domain is proportional to the intensity of its Bragg reflection. These effects can be seen in the X-ray data of Fig.~\ref{4-bragg}, especially the $T$=132~K panels.}
\label{5-schematics}%
\end{figure}

\section{Results}

\subsection{Polarized Microscopy}

The unit cell in BaFe$_2$As$_2$ doubles in size and rotates by 45 degrees upon cooling through the tetragonal to orthorhombic transition. This leads to the formation of the domain walls at 45 degrees with respect to the sample edges (see bottom left panel of Fig.~\ref{1-detwin}). The orthorhombic $\bf{a_o}$ and $\bf{b_o}$ axes inside the domains are at 45 degrees to the twin boundaries (see Ref. ~\onlinecite{domains}). Therefore the highest contrast of domain imaging is achieved when the sample is aligned with a long [110] tetragonal direction at 45$^o$ to the polarization direction of the linearly polarized light (parallel and perpendicular to the orthorhombic $\bf{a_o}$ in different domains.)

The optical contrast of the domains is determined by the anisotropy of the bi-reflectance and is proportional to the orthorhombic distortion. It increases with decreasing temperature, and is weaker in Ba122 than in Ca122 \cite{domains}. Simultaneously, due to dispersion of the bi-reflectance, initially white incident light on reflection acquires color depending on the orientation of the orthorhombic axes with respect to the polarization direction of the incident light. This results in different colors of structural domains as can be seen in the bottom left panel of Fig.~\ref{1-detwin}. Therefore, the color of the image contains information about the orthorhombic distortion, albeit in arbitrary units, and can be used for the analysis of its temperature dependence even in the detwinned state below the structural transition or in the strained state above the transition.

Figure~\ref{f2-RGB} shows images from the strained region B of Fig.~\ref{1-detwin}, below (130~K, left top panel) and above (160~K, left bottom) the transition. This region is completely detwinned by strain. The red square shows the small clean area of the sample, where the color of the image was analyzed numerically. The right panels in Fig.~\ref{f2-RGB} show red-green-blue (RGB) histograms of that area. The images were taken every 5 K from 80 to 260 K. Fig.~\ref{3-color} shows the difference between the intensities of the blue and green channels, indicating that the structural distortion does not vanish abruptly at the transition but remains notable up to 200~K.

For reference we show the results of the equivalent analysis in the parent compound Ca122. Here the transition is strongly first order, and the data show no tail above the transition.

\subsection{X-ray diffraction}

X-ray analysis was done in both the unstrained (area A) and strained (area B) parts of the same crystal, as shown in Fig.~\ref{1-detwin}. In the unstrained region, the tetragonal (220) Bragg peak splits below $T_{TO}$ into four peaks, each representing a distinct orthorhombic domain. The four orthorhombic reflections merge back into a single tetragonal Bragg peak on warming above $T_{TO}$ (see Fig. 4).

The integrated intensities of the reflections at 6~K allow us to determine the relative population of orthorhombic domains. In the unstrained area, the population of the four domains ranges 19 to 31 \% of the total peak intensity, characteristic of a near random distrobution. In the strained region, the domain whose $\bf{a_O}$-axis lies along the direction of the strain accounts for nearly 90 \% of the probed sample volume.

These effects are schematically described in Fig.~\ref{5-schematics}. The separation between the Bragg peaks resulting from the orthorhombic O1 and O2 domains is fixed because the relative angle between their twinning planes is fixed. The same is true for the peaks from the O3 and O4 domains. However there exists no such rule for the separation between the O2 and O4 peaks because the angle between their twinning planes is determined by their relative domain populations. In a sample with perfectly equal domain populations the four Bragg peaks would produce a square with each reflection having equal intensity. As the relative population of one domain orientation grows, the angle between the twinning planes of the O2 and O4 peaks becomes smaller and consequently the separation of the their Bragg peaks diminishes. This behavior is readily seen in the X-ray data in Fig.~\ref{4-bragg}. The unstrained region of the crystal manifests relatively similar populations of each domain orientation and produces a pattern not quite square but slightly trapezoidal below the transition temperature. By contrast in the strained region of the crystal, where the dominant domain orientation represents nearly 90\% of the sample volume, the O2 and O4 peaks are no longer distinguishable as two separate reflections.

\begin{figure}[tb]
\begin{center}
\includegraphics[width=0.85\linewidth]{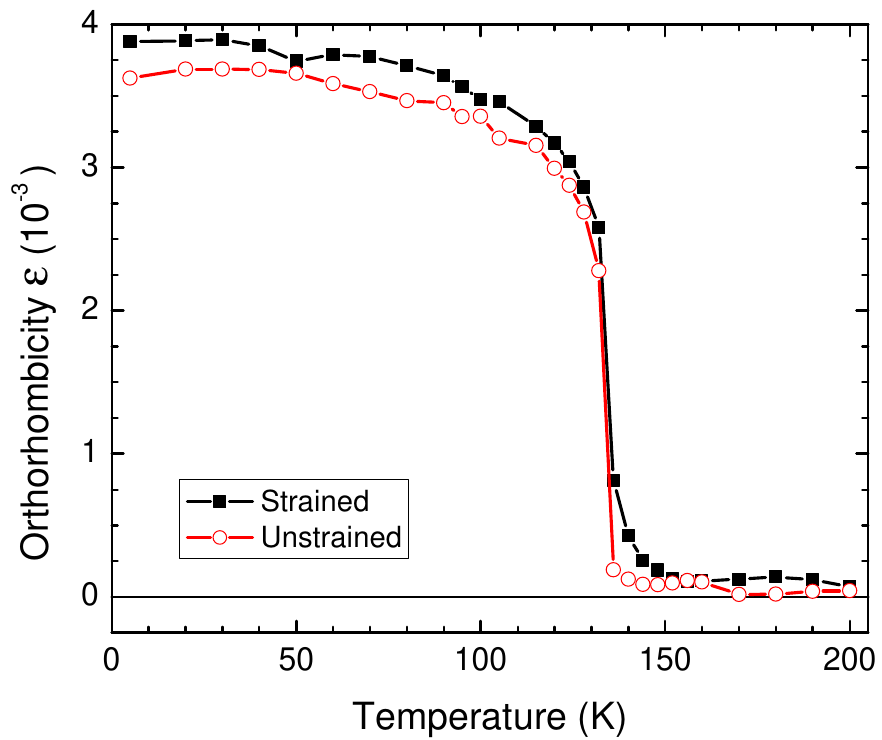}
\end{center}
\caption{Orthorhombic distortion, $\epsilon= \frac{(a_O-b_O)}{(a_O+b_O)}$, vs. temperature in strained and unstrained parts of the sample. The strain notably increases the orthorhombic distortion below the transition, and induces a ``tail'' of orthorhombic distortion above the sharp drop at 135~K. }%
\label{5-delta}%
\end{figure}

The temperature evolution of the orthorhombic distortion,  $\epsilon \equiv \frac{(a_O-b_O)}{(a_O+b_O)}$, can be clearly seen as an increased splitting distance between the orthorhombic reflections at 6 K as compared to the splitting at 132 K (see Fig.~\ref{4-bragg}), and is shown in Fig.~\ref{5-delta}. Application of strain notably increases $\epsilon$ below the transition, and most importantly, a ``tail'' of the orthorhombic distortion can be tracked to at least 150~K, well above 135~K where the order parameter, $\epsilon$, shows a sharp drop.

\subsection{Resistivity}

\begin{figure}[tb]
\begin{center}
\includegraphics[width=1\linewidth]{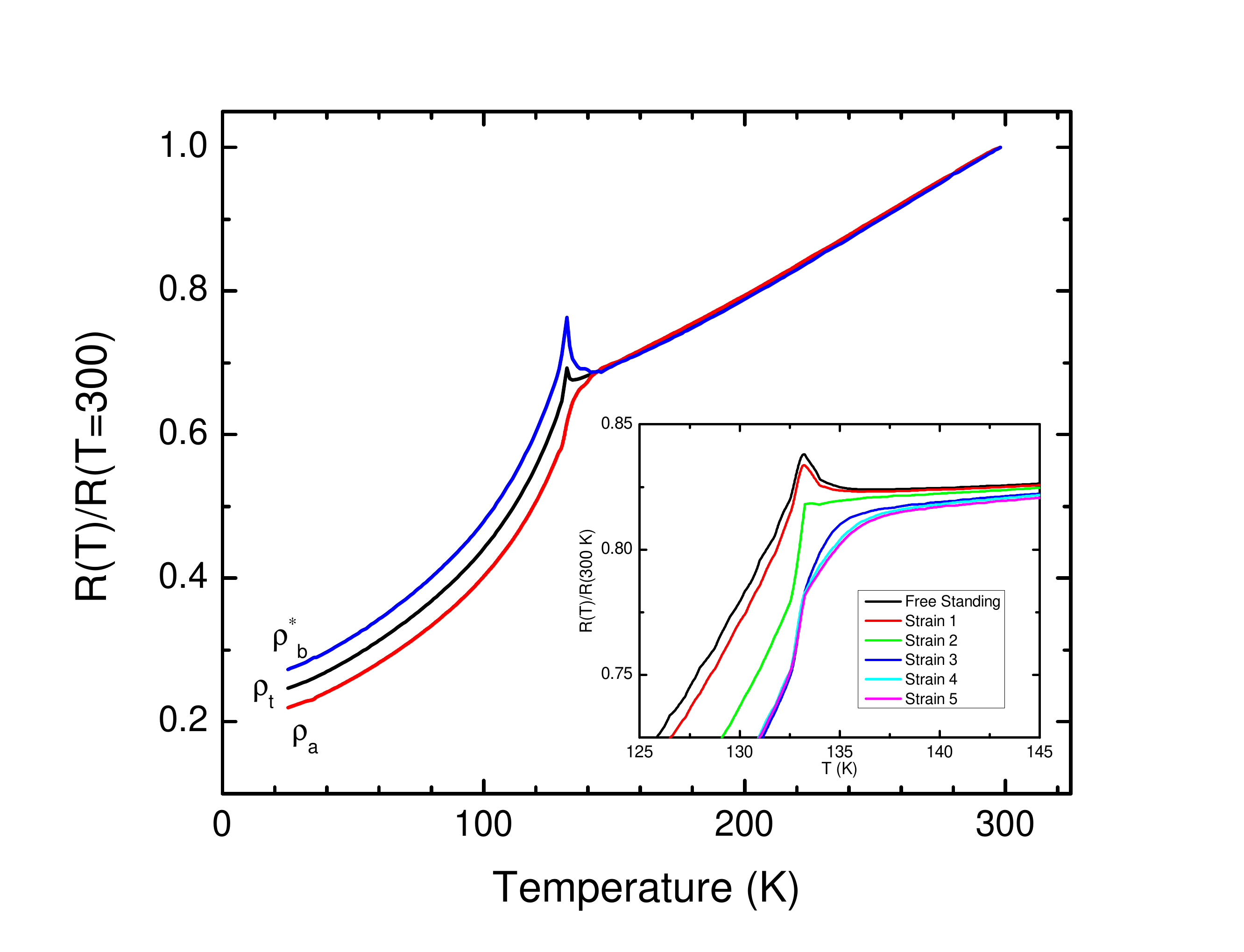}
\end{center}
\caption{(Color Online) Temperature-dependent normalized resistivity, $\rho_a$(T)/$\rho_a$(300 K) of the BaFe$_2$As$_2$ sample in the free-standing, $\rho _t$ (black curve), and strain-detwinned, $\rho _a$ (red curve) regions of the same sample used for X-ray measurements in Figs~\ref{4-bragg},~\ref{5-delta}, and~\ref{8-SrBa}. The third (blue) curve shows $\rho^* _b$, calculated as $\rho^*_b= 2*\rho_t -\rho_a$.  The anisotropy can be seen for all temperatures below the transition, and a slight anisotropy can be found above the transition. Inset: Progression of the effect of increasing strain on the resistivity ($\rho_a$ in the detwinned state). The black curve represents a free standing crystal. Tensile stress incrementally increases until reaching approximately 20 MPa for strain 5, see text for details. Strain 2 is sufficient to detwin the sample, revealing a sharp drop in resistivity at the transition. On further strain increase the jump rounds and its onset shifts up in temperature. }%
\label{6-resistivity}%
\end{figure}

Figure~\ref{6-resistivity} shows the normalized temperature-dependent resistivity in the twinned, $\rho _t$, and strain-detwinned, $\rho _a$, states of the same sample as measured by $x$-ray diffraction (Fig~\ref{4-bragg},~\ref{5-delta}). The third curve was calculated assuming that $\rho _t$ represents an equal mixture of $\rho _a$ and $\rho _b$, $\rho^*_b \equiv 2\rho_t-\rho_a$.

After sufficient stress was applied to detwin the crystals (the sample in Fig. 7 was nearly completely detwinned after the second strain increment, Strain 2, as determined by polarized optical imaging), we performed a careful study of the effect of additional stress on the resistivity anisotropy. The stress, whose magnitude at the highest level is estimated to be in the 20 MPa range, increases the onset temperature of the resistivity anisotropy. However the most dramatic effects of the resistivity change are clearly around the point where the strain is sufficient to detwin the crystal.

\section{Discussion}

\subsection{Strain-induced Anisotropy}

\begin{figure}[tb]
\begin{center}
\includegraphics[width=0.85\linewidth]{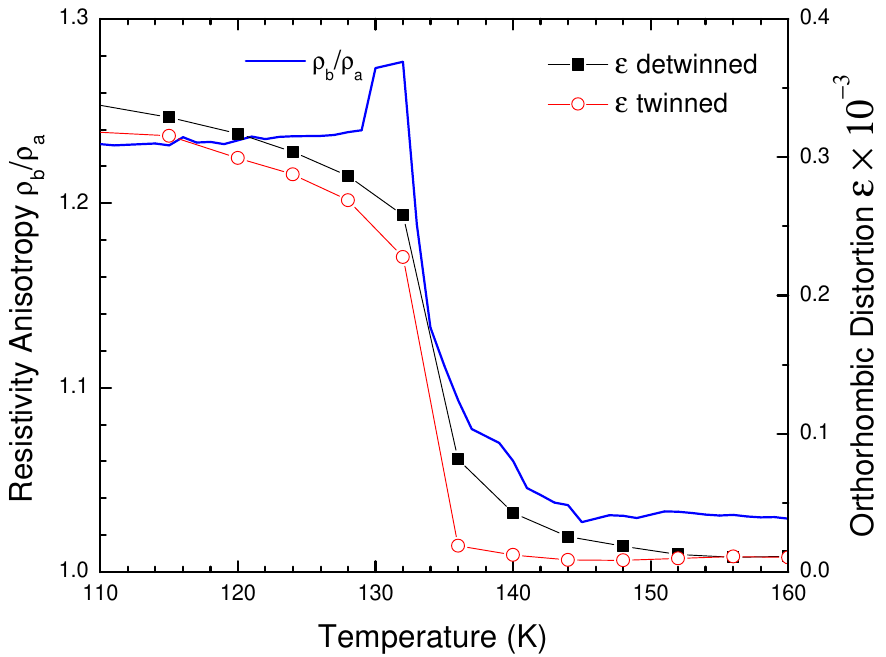}
\end{center}
\caption{(Color Online) Comparison of the temperature-dependent resistivity anisotropy, $\frac{\rho^* _b}{\rho_a}$,  and the orthorhombic distortion,  $\epsilon= \frac{(a_O-b_O)}{(a_O+b_O)}$ in the temperature range close to $T_{TO}$. Both quantities show a pronounced ``tail'' above a sharp drop in the order parameter at 135~K, revealing that the anisotropy is directly related to strain.}
\label{7-comparison}%
\end{figure}

Figure~\ref{7-comparison} shows a direct comparison of the temperature-dependent degree of the orthorhombic distortion, $\epsilon (T)$, and of the resistivity anisotropy, $\rho _b / \rho _a$, in the same sample of BaFe$_2$As$_2$ under identical strain conditions. The two quantities reveal a clear correlation. Both show a rapid rise below approximately 135~K with decreasing temperature. In addition both $\epsilon$ and $\rho_b/\rho_a$ show a clear ``tail'' above 135~K, in agreement with the color analysis discussed above. In strain-free samples of Ba122, the tetragonal-to-orthorhombic structural transition at $T_{TO}$ is of the second order and precedes a strongly first-order magnetic transition at $T_N \le T_{TO}$ \cite{Kreyssigsplit}. It is then natural to assign the rapid increase of the anisotropy below 135~K to a magnetic transition, while the ``tail'' above 135~K correlates with the orthorhombic distortion. The exact meaning of the structural transition in the presence of the strain field becomes unclear, as the order parameter varies smoothly with temperature. Therefore from this direct comparison we conclude that externally applied strain is the cause of the structural and transport anisotropy above $T_{TO}$.

\subsection{Comparison of the effect of strain on first and second order transition: BaFe$_2$As$_2$ vs SrFe$_2$As$_2$}

\begin{figure}[tb]
\begin{center}
\includegraphics[width=0.85\linewidth]{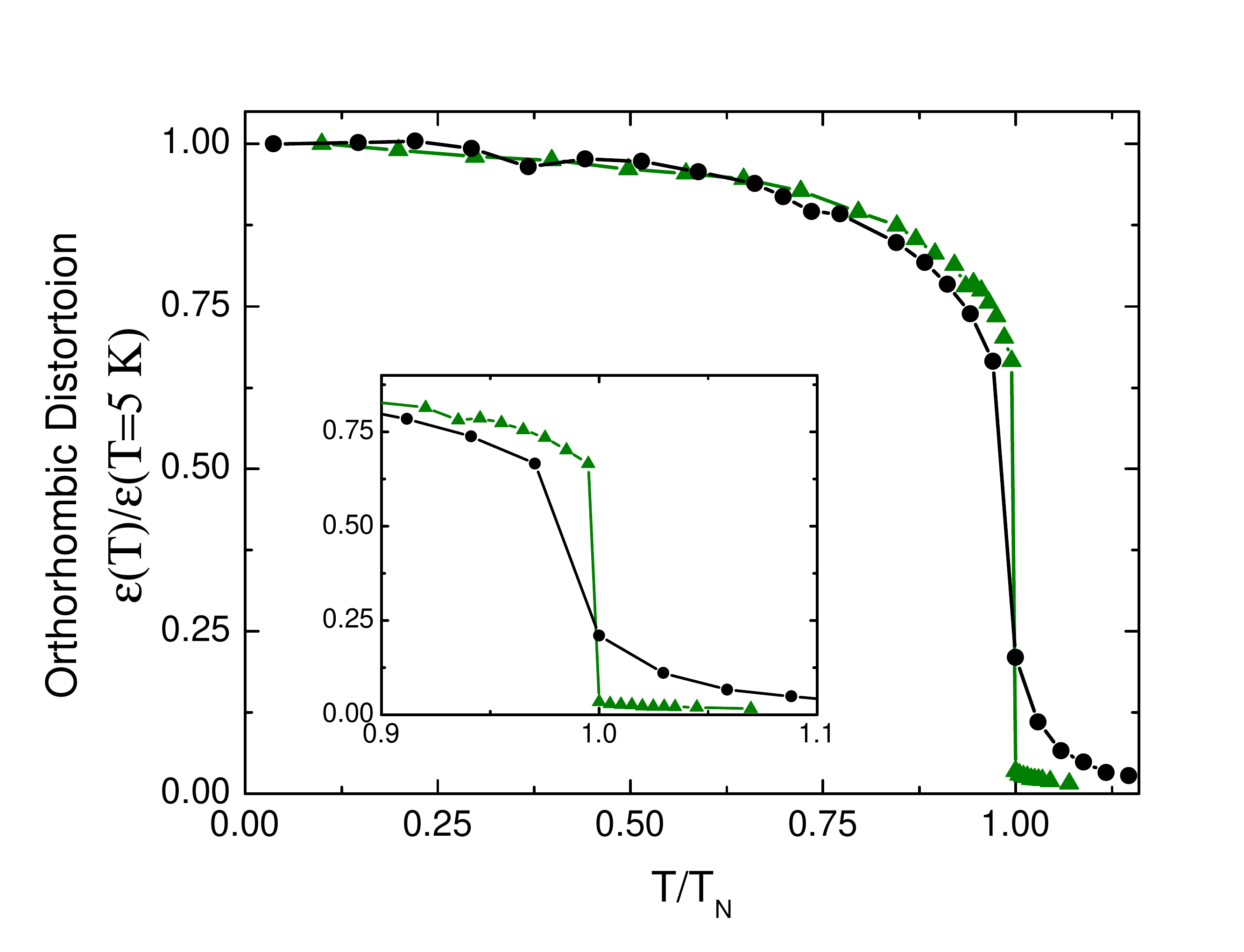}
\end{center}
\caption{(Color Online) Comparison of the temperature-dependent orthorhombic distortions, $\epsilon=\frac{a_O-b_O}{a_O+b_O}$,  in strain detwinned areas of SrFe$_2$As$_2$, Ref.~\onlinecite{detwinning2}, and BaFe$_2$As$_2$. The data are presented vs. normalized temperature $T/T_N$. A pronounced ``tail'' above $T_N$ in BaFe$_2$As$_2$ is caused by an anomalously strong susceptibility of the lattice to strain.
}%
\label{8-SrBa}%
\end{figure}

In Fig.~\ref{8-SrBa} we compare the temperature dependent orthorhombic distortions, $\epsilon$, for two strained A122 compounds each with a very different character of the transition: strongly first-order in Sr122 and second order in Ba122. The data are plotted on a normalized temperature scale, $T/T_N$. As is clear from the Figure~\ref{8-SrBa}, the ``tail'' of the anisotropy above the transition is virtually absent in Sr122, whereas it is quite noticeable in Ba122. In the next section we apply Ginzburg-Landau - type theory to model the effect of strain on the resistivity anisotropy considering first and second order transitions in the strain field.

\subsection{ Phenomenological model of the effect of the uniaxial strain }

Regardless of which electronic degree of freedom $\varphi$ is responsible for the electronic anisotropy, it should be proportional to the orthorhombic distortion, since both break the tetragonal symmetry of the lattice close to $T_s$. For example, $\varphi$ can be associated with magnetic fluctuations \cite{Rafael1,Rafael1}. By symmetry, $\phi$ and $\epsilon$ are bilinearly coupled in the free energy expansion, ie. they give rise to the term $\phi$$\epsilon$. Since the external strain $\sigma$ also couples bilinearly to the orthorhombic distortion $\epsilon=(a_O-b_O)/(a_O+b_O)$, it has an effect on $\phi$ similar to that of a magnetic field $h$ on Ising ferromagnets.

\begin{figure}[tb]
\begin{center}
\includegraphics[width=0.85\linewidth]{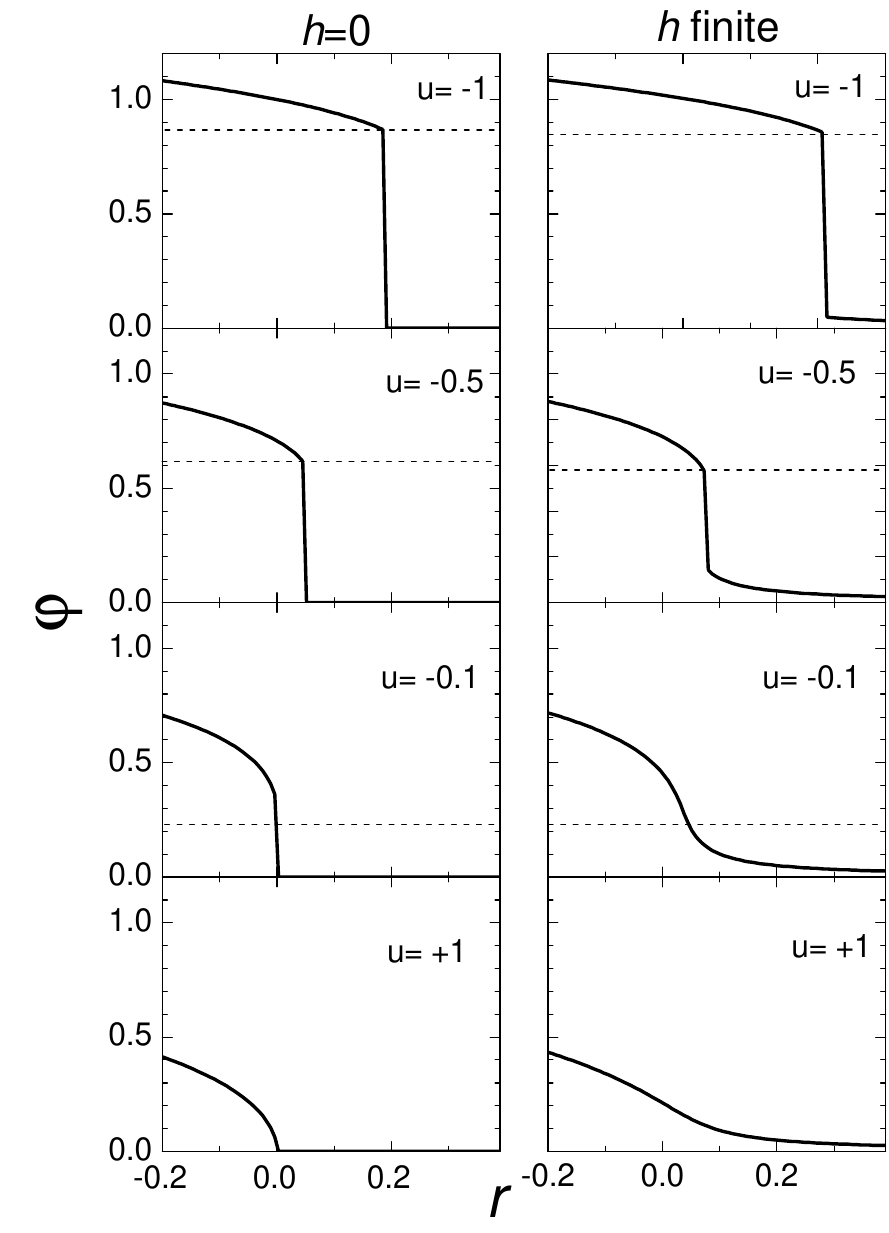}
\end{center}
\caption{(Color Online) Evolution of the anisotropy parameter $\varphi\propto\rho_{b}-\rho_{a}$ vs. temperature parameter $r$,
 $r\propto T-T_{s}^{0}$. Left column of panels is for
$h=0$, right column is for $h=0.01w$. The bottom pair of panels shows a second-order transition for $u$=$w$=1, the other
pairs of the panels show first-order
transitions for u=-w, u=-0.5w and u=-0.1w (top to bottom). The dashed lines show
the size of the jump $\Delta\varphi$ in the absence of an external
field.}%
\label{Rafael1}%
\end{figure}

In order to compare the effect of a finite $h$ on the second-order and the first-order structural phase transitions, we consider the phenomenological free energy:

\begin{equation}
F=\frac{r}{2}\varphi^{2}+\frac{u}{4}\varphi^{4}+\frac{w}{6}\varphi^{6}-h\:\varphi\label{free_energy}\end{equation}

\noindent
with temperature parameter $r\propto T-T_{s}^{0}$, where $T_{s}^{0}$ is the mean-field structural transition temperature. Here $u$ and $w$ are phenomenological parameters of Ginzburg-Landau theory describing the
phase transition. To ensure the stability of the free-energy expansion, $w$ has to be positive. If $u>0$ as well, we have a second-order transition. If $u<0$, we have a first-order transition. In this case, the ratio $-u/w$ determines how strong the first-order transition is, i.e. what is the magnitude of the jump of the order parameter. For $u>0$, we have a second-order phase transition at $r=0$ for $h=0$. The effect of a small but finite $h$ is to extend the region of finite $\varphi$ asymptotically to $r\rightarrow\infty$, giving rise to a ``tail'' in the plot of $\varphi$ as function of temperature (see Fig.~\ref{Rafael1}). Formally, there is no strict $T_{s}$, although experimentally there will be a temperature above which the distortion anisotropy is too small to be detected. Notice that, at $T_{s}^{0}$, the value of $\varphi$ scales with the applied field according to $\varphi\sim h^{1/\delta}$, where $\delta=3$ is the mean-field critical exponent.

Let us consider $u<0$, which gives rise to a first-order phase transition. As usual for first-order phase transitions, there is a coexistence region where the states with $\varphi=0$ and $\varphi\neq0$ are both local minima of the free energy. If we consider an adiabatic change of temperature, such that the system always chooses the global minimum,  from the minimization of Eq. \ref{free_energy} it is straightforward to find that the transition takes place above $T_{s}^{0}$, at $\frac{r}{w}=\frac{9}{48}\left(\frac{u}{w}\right)^{2}$. We also find that the jump in the order parameter is $\Delta\varphi=\sqrt{\frac{-3u}{4w}}$.

Therefore, the ratio $\left|u\right|/w$ controls the strength of the first-order transition. The effect of a finite field on the jump $\Delta\varphi$ will depend on the value of the ratio $h/\left|u\right|$. In Figure~\ref{Rafael1}, we plot the temperature evolution of $\varphi$ for different values of $\left|u\right|/w=\left\{ 1,\:0.5,\:0.1\right\} $, keeping $h/w=0.01$ constant. The dashed line shows the magnitude of the jump $\Delta\varphi$ for $h=0$. Notice that when the first-order transition is stronger, the jump is barely affected by the finite field.
In particular, above the temperature where the jump takes place, the
order parameter is never zero but is always very small, giving rise
to a rather small ``tail''. On the other hand, when the first-order
transition is weaker, the same field can completely smear out the jump.
This gives rise to a noticeable and continuous ``tail'', and therefore to a second-order transition \cite{Millis}.

This analysis suggests that the anisotropy above the second order transition originates from the fact that the orthorhombic transition is actually not strictly defined in the strained (and thus orthorhombically distorted) tetragonal phase under uniaxial stress. On the other hand it suggests that the susceptibility to stress is notably enhanced in case of a weak second order transition character.

\section{Conclusions}

Systematic characterization of the effect of permanently applied stress on the properties of BaFe$_2$As$_2$ using $x$-ray, polarized optics and electrical resistivity measurements suggest that the applied stress is actually the cause of the resistivity anisotropy in the nominally tetragonal phase. Thus the
resistivity anisotropy ``tail'' above the temperature of the structural transition is solely due to the effect of the uniaxial strain applied  to detwin the samples. The difference between $A$Fe$_2$As$_2$ compounds with various alkali earth metals is determined by the character and the strength (order parameter jump) of the structural transition. These conclusions are supported by a phenomenological model of the effect of the uniaxial strain on the structural transition, similar to the effect of a magnetic field on Ising ferromagnets. It is interesting to notice that this strong susceptibility to the effect of strain is also found in the orbital splitting in ARPES measurements \cite{Shen,Korean}, and in the temperature dependence of the shear modulus in ultrasonic experiments \cite{Rafael2}.

\subsection{Acknowledgement}

We thank D. Robinson for the excellent support of the high--energy x-ray scattering study. Use of the Advanced Photon Source was supported by the U. S. Department of Energy, Office of Science, under Contract No. DE-AC02-06CH11357. Work at the Ames Laboratory was supported by the U.S. Department of Energy, Office of Basic Energy Sciences, Division of Materials Sciences and Engineering under contract No. DE-AC02-07CH11358.


\end{document}